%% file: main.tex
\documentclass[conference]{IEEEtran}
\IEEEoverridecommandlockouts

\usepackage{cite}
\usepackage{amsmath,amssymb,amsfonts}
\usepackage{algorithmic}
\usepackage{graphicx}
\usepackage{textcomp}
\usepackage{xcolor}
\usepackage{siunitx}

\usepackage{threeparttable}
\usepackage{booktabs}

\usepackage{acronym}
\usepackage{float}
\usepackage{multirow}
\usepackage{soul}
\usepackage{comment}
\usepackage{makecell} 

\usepackage{xcolor}

\usepackage{array}
\newcolumntype{P}[1]{>{\centering\arraybackslash}p{#1}}
\newcolumntype{M}[1]{>{\centering\arraybackslash}m{#1}}

\def\BibTeX{{\rm B\kern-.05em{\sc i\kern-.025em b}\kern-.08em
    T\kern-.1667em\lower.7ex\hbox{E}\kern-.125emX}}

  \makeatletter
  \let\ps@IEEEtitlepagestyle\ps@mahmood
  \makeatother
\makeatletter
\def\ps@IEEEtitlepagestyle{%
  \def\@oddfoot{\mycopyrightnotice}%
  \def\@oddhead{\hbox{}\@IEEEheaderstyle\leftmark\hfil\thepage}\relax
  \def\@evenhead{\@IEEEheaderstyle\thepage\hfil\leftmark\hbox{}}\relax
  \def\@evenfoot{}%
}
\def\mycopyrightnotice{%
  \begin{minipage}{\textwidth}
  \centering \scriptsize
  \copyright 2023 IEEE.  Personal use of this material is permitted.  Permission from IEEE must be obtained for all other uses, in any current or future media, including reprinting/republishing this material for advertising or promotional purposes, creating new collective works, for resale or redistribution to servers or lists, or reuse of any copyrighted component of this work in other works.
  \end{minipage}
}

\begin{document}

\acrodef{HR}{Heart Rate}
\acrodef{BLE}{Bluetooth Low Energy}
\acrodef{DSP}{Digital Signal Processing}
\acrodef{NN}{Neural Network}
\acrodef{EEG}{Electroencephalography}
\acrodef{FFC}{Flexible Flat Cable}
\acrodef{ULP}{Ultra Low Power}
\acrodef{IMU}{Inertial Measurement Unit}
\acrodef{FR}{frame rate}
\acrodef{PMIC}{Power Management Integrated Circuit}
\acrodef{WULPUS}{Wearable Ultra Low-Power Ultrasound}
\acrodef{US}{Ultrasound}
\acrodef{ULP}{ultra low-power}
\acrodef{AFE}{analog front-end}
\acrodef{RF}{radio frequency}
\acrodef{ML}{machine Learning}
\acrodef{FPGA}{Field-Programmable Gate Array}
\acrodef{LVDS}{low-voltage differential signalling}
\acrodef{SPI}{serial peripheral interface}
\acrodef{MCU}{microcontroller unit}
\acrodef{ADC}{analog to digital Converter}
\acrodef{HDL}{hardware description language}
\acrodef{DMA}{direct memory access}
\acrodef{HMI}{Human-Machine Interface}
\acrodef{HD}{hardware design}
\acrodef{AD}{algorithm development}
\acrodef{FPS}{frames per second}
\acrodef{DAS}{delay-and-sum}
\acrodef{TGC}{time-gain compensation}
\acrodef{FP}{floating point}
\acrodef{BW}{bandwidth}

\acrodef{NE16}{Neural Engine 16}
\acrodef{CNN}{convolutional neural network}
\acrodef{RNN}{recurrent neural network}
\acrodef{FFT}{Fast Fourier Transform}
\acrodef{SSVEP}{Steady State Visually Evoked Potential}
\acrodef{BCI}{Brain-Computer Interface}
\acrodef{PULP}{Parallel Ultra Low Power}
\acrodef{SoC}{System on Chip}
\acrodef{BLE}{Bluetooth Low Energy}
\acrodef{EEG}{electroencephalography}
\acrodef{EMG}{electromyography}
\acrodef{sEMG}{Surface electromyography}
\acrodef{ECG}{electrocardiogram}
\acrodef{PPG}{photoplethysmogram}
\acrodef{GPIO}{General Purpose Input/Output}
\acrodef{PGA}{Programmable Gain Amplifier}
\acrodef{HV}{high voltage}
\acrodef{WL}{Waveform Length}
\acrodef{BP}{Band-pass}
\acrodef{SoA}{state-of-the-art}


\newcommand{\victor}[1]{\textcolor{orange}{#1}}

\title{A Wearable Ultra-Low-Power sEMG-triggered Ultrasound System for Long-term\newline Muscle Activity Monitoring
\thanks{The authors acknowledge support from the Swiss National Science Foundation (Project PEDESITE, grant agreement 193813) and from ETH Z{\"u}rich (project ListenToLight, grant agreement ETH-C-01 21-2). This work was also partially supported by the ETH Future Computing Laboratory (EFCL).}

}

\author{
  \IEEEauthorblockN{
    Sebastian Frey\IEEEauthorrefmark{1},
    Victor Kartsch\IEEEauthorrefmark{1},
    Christoph Leitner\IEEEauthorrefmark{1},
    Andrea Cossettini\IEEEauthorrefmark{1},
    Sergei Vostrikov\IEEEauthorrefmark{1},\\
    Simone Benatti\IEEEauthorrefmark{3}\IEEEauthorrefmark{2},
    Luca Benini \IEEEauthorrefmark{1}\IEEEauthorrefmark{2}\\
    }
    
    \vspace{0.2cm}

  \IEEEauthorblockA{\IEEEauthorrefmark{1}Integrated Systems Laboratory, ETH Z{\"u}rich, Z{\"u}rich, Switzerland}
  \IEEEauthorblockA{\IEEEauthorrefmark{2}DEI, University of Bologna, Bologna, Italy}
  \IEEEauthorblockA{\IEEEauthorrefmark{3}DISMI, University of Modena and Reggio Emilia, Reggio Emilia, Italy}
}

\maketitle
\IEEEpubidadjcol

\begin{abstract}
\input{sections/abstract}
\end{abstract}


\input{sections/intro}
\input{sections/SYSTEM_v2}
\input{sections/experimental_verif_v2}

\input{sections/conclusion}

\section*{Acknowledgment}
\vspace{-0.1cm}
We thank A. Blanco Fontao and H. Gisler (ETH Zürich) for technical support.

\bibliographystyle{IEEEtran}
\bibliography{bibliography.bib}
\end{document}

%% file: sections/abstract.tex
\ac{sEMG} is a well-established approach to monitor muscular activity on wearable and resource-constrained devices. However, when measuring deeper muscles, its low signal-to-noise ratio (SNR), high signal attenuation, and crosstalk degrade sensing performance. \ac{US} complements sEMG effectively with its higher SNR at high penetration depths. In fact, combining \ac{US} and sEMG improves the accuracy of muscle dynamic assessment, compared to using only one modality. However, the power envelope of \ac{US} hardware is considerably higher than that of sEMG, thus inflating energy consumption and reducing the battery life. This work proposes a wearable solution that integrates both modalities and utilizes an EMG-driven wake-up approach to achieve ultra-low power consumption as needed for wearable long-term monitoring. We integrate two wearable \ac{SoA} \ac{US} and ExG biosignal acquisition devices to acquire time-synchronized measurements of the short head of the biceps. To minimize power consumption, the US probe is kept in a sleep state when there is no muscle activity. sEMG data are processed on the probe (filtering, envelope extraction and thresholding) to identify muscle activity and generate a trigger to wake-up the \ac{US} counterpart. The \ac{US} acquisition starts before muscle fascicles displacement thanks to a triggering time faster than the electromechanical delay (30-100 ms) between the neuromuscular junction stimulation and the muscle contraction. Assuming a muscle contraction of 200~ms at a contraction rate of 1~Hz, the proposed approach enables more than 59\% energy saving (with a full-system average power consumption of 12.2~mW) as compared to operating both sEMG and \ac{US} continuously.

%% file: sections/intro.tex
\vspace{-0.1cm}
\section{Introduction}
\vspace{-0.1cm}

Surface electromyography (sEMG) is a well-established approach for determining muscular activity on wearable and resource-constrained devices \cite{cerone2019modular}. Its popular applications include the detection of muscle activations \cite{j:DeLuca1997} and their integration into human-machine interfaces (HMI) \cite{j:Benatti2015} or prosthetic \cite{w:ottobock} control systems. In sEMG, activation signals of muscles are acquired from the skin surface \cite{reaz2006techniques}. These collected sEMG signals reflect a total activation sum of all underlying muscle tissue \cite{athavale2017biosignal}. As a consequence, an unambiguous correlation between surface-collected signals and the contribution of specific muscles is not always possible.  Moreover, sEMG suffers from a low signal-to-noise ratio (SNR) and interference from neighbouring muscles \cite{CLANCY20021}. Thus, complementing sEMG with alternative sensing modalities to enhance the quality and reliability of the measured signals is highly desired.

An alternative modality to analyze musculoskeletal activity is \ac{US} \cite{j:vanhooren2019, b:Lichtwark2017}. \ac{US} has several advantages over sEMG, including a higher SNR at high penetration depths, lower susceptibility to noise and interference, and higher spatial resolution, as demonstrated by a number of different applications, such as elastography \cite{Tanter2002UltrafastElastography}, motor endplates detection \cite{c:Leitner2020:1}, strain analyses \cite{c:Leitner2020:2}, and even gestures \cite{Vostrikov2023hand} or finger motion recognition \cite{yang2018towards}. With US, however, only the mechanical response to an electrically triggered voluntary contraction can be observed. Therefore, lacking information about the electro-chemically induced neuronal activity. In this context, the combination of sEMG and \ac{US} appears as a promising approach for achieving higher accuracy in the assessment of muscle dynamics \cite{botter2013novel}.

To enable continuous, long-term monitoring, there is a need for wearable solutions that are wireless, unobtrusive, and low power. Among these requirements, power consumption appears as the main challenge for the successful integration of sEMG and \ac{US} into wearables.
In fact, \ac{US} hardware is more power-hungry than that of sEMG\footnote{considering two \ac{SoA} wearable ExG and \ac{US} platforms \cite{frey2023biogap}\cite{frey2022wulpus}, the generation of high-voltage pulses and analog-to-digital conversion for a single-channel \ac{US} consumes nearly twice the power of the analog-to-digital conversion of a single-channel ExG}, and implementing concurrent, continuous sEMG and \ac{US} measurements would use too much energy and drain the battery of wearable sensor systems. Therefore, to achieve low power consumption, a wearable solution that combines both modalities should use physiology knowledge to activate \ac{US} measurements only when the signal is meaningful.

In this paper, we propose a wearable sensor system that combines sEMG and \ac{US} for measuring muscle activity. The system is based on two \ac{SoA} wearable probes for ExG \cite{frey2023biogap} and \ac{US} \cite{frey2022wulpus}. The \ac{US} sub-system is kept in a sleep state when there is no muscle activity. The sEMG signal is acquired continuously, and we use an EMG-driven wake-up approach to identify muscle activity (the envelope of sEMG signals is extracted and compared to a threshold), generating a trigger to wake-up the \ac{US} counterpart only when significant muscle activity is detected. The triggering is faster than the electromechanical delay (\SI{30}{-}\SI{100}{ms}) between the neuromuscular junction stimulation and the muscle contraction, thereby waking up the \ac{US} sub-system before the actual displacement of muscle fascicles starts.
We focus on the use case of biceps muscle contractions. By implementing the proposed wake-up approach directly on-probe, we demonstrate that the proposed methodology allows to considerably reduce the \ac{US} hardware's power consumption: assuming contractions of \SI{200}{ms} at a contraction rate of \SI{1}{Hz}, the total system power is only \SI{12.2}{mW}, $2.4\times$ lower compared to operating both sEMG and \ac{US} continuously leading to a battery life increase of more than 2 days with a \SI{320}{mAh} battery.

%% file: sections/SYSTEM_v2.tex
\section{Material and Methods}
\label{sec:methods}

\begin{figure}[t]
\centerline{\includegraphics[width=1\columnwidth,trim={0.8cm 0 0.2cm 0},clip]{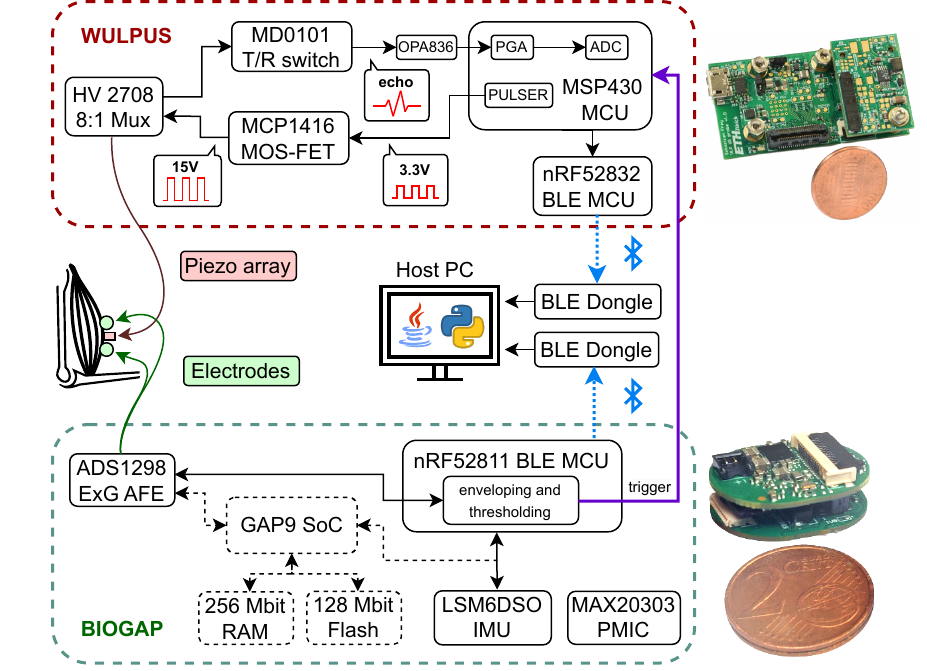}}
\vspace{-0.3cm}
\caption{System diagram of the combined US-ExG platform.}
\vspace{-0.4cm}
\label{fig:prototype}
\end{figure}

\subsection{System description}
The framework proposed in this study combines two \ac{SoA} ExG and US acquisition platforms, namely BioGAP \cite{frey2023biogap} and WULPUS \cite{frey2022wulpus}.

\subsubsection{sEMG data acquisition subsystem.}
BioGAP (see Fig.~\ref{fig:prototype}, bottom box) is a modular biosignal acquisition and processing platform for ExG biosignals. BioGAP integrates an nRF52811 \ac{SoC} (Nordic Semiconductor) for \ac{BLE} connectivity and measurement control, a GAP9 \ac{PULP} processor (GreenWaves Technologies) for online data processing, an accelerometer for movement sensing, and a dedicated power management circuit. A specialized \ac{AFE} (ADS1298, Texas Instruments) is employed for biopotential measurements and is configured as detailed in Tab.~\ref{table:meas_config_ADS1298}. BioGAP supports both active and passive electrodes, providing flexibility for various sEMG recording setups.
In this work, the BioGAP is configured in a single-channel EMG mode with active electrodes to detect the contraction of the short head of the biceps.

\input{tables/ADS1298_configuration}

\subsubsection{US data acquisition subsystem.}
WULPUS (see Fig.~\ref{fig:prototype}, top box) is an ultra-low-power 8-channel (time-multiplexed) wearable \ac{US} probe, enabling raw data acquisition and wireless transmission to a host PC. The probe utilizes the MSP430 \ac{MCU} (Texas Instruments), specifically designed for \ac{US} applications, coupled to a nRF52832 \ac{SoC} (Nordic Semiconductor) for \ac{BLE} connectivity. The operation of the system can be summarized as follows. First, high-frequency \SI{3.3}{V} unipolar pulses are generated by the MSP430 \ac{US} \ac{MCU}. Subsequently, these pulses are amplified by a gate driver and forwarded to a transducer array via an 8:1 \ac{HV} multiplexer. After the pulse transmission phase, the multiplexer is switched to receive mode and the backscattered US signals travel back through a transmit/receive switch and an amplifier to the MSP430 \ac{MCU}. Here, the US signal is sampled at a rate of \SI{8}{MHz}, and the data is transmitted to the nRF52832 SoC using DMA-controlled SPI. Finally, the nRF52832 \ac{MCU} sends the data via \ac{BLE} to a PC for data logging and visualization.

\subsubsection{System integration and triggering.}
Time synchronization and triggering across the two subsystems is enabled by a direct connection between the nRF52811 \ac{MCU} of BioGAP and the MSP430 \ac{MCU} of WULPUS (purple arrow in Fig.~\ref{fig:prototype}). 

\subsection{Experimental Measurement Setup}

For the experimental validation of our device and algorithm we employed measurements on a biceps dynamometer. The measurement system and placement of the study participant on the testbench was as described in \cite{c:leitner2021}. We acquired time-synchronized data (\ac{sEMG}, US and force) of isometric voluntary contractions on the biceps brachii short head. 

\subsubsection{Placement of sEMG electrodes and US transducer}
\begin{figure}[tb]
\centerline{\includegraphics[width=0.8\columnwidth]{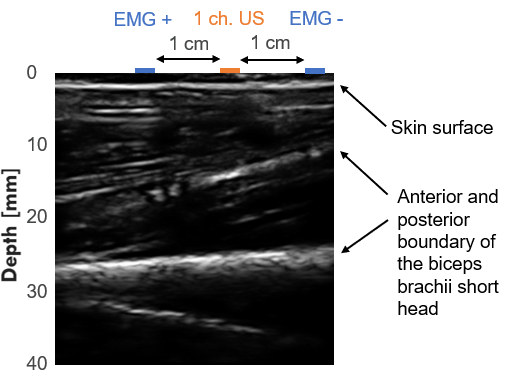}}
\vspace{-0.3cm}
\caption{Placement of the EMG electrodes and the US transducer.}
\vspace{-0.4cm}
\label{fig:placement_sensor}
\end{figure}

\ac{sEMG} signals were acquired with one differential channel using two active, wet electrodes placed on the biceps brachii according to the SENIAM guidelines \cite{seniam_guidelines}. The interelectrode distance was \SI{3}{cm}. The bias electrode was attached to the elbow.
US measurements are based on a linear array transducer with 32 channels and a central frequency of \SI{2.25}{MHz} (LA-32, Vermon), which offers a favourable trade-off between penetration depth and axial resolution. The transducer was placed between the \ac{sEMG} electrodes and fixed on biceps brachii using an elastic band (see Fig.~\ref{fig:prototype}). Short cables were used to interconnect the transducer with WULPUS. The probe was configured to drive eight central channels simultaneously in a plane-wave mode (for increased transmit energy) and to receive the echo signal from a single (the centre-most) channel.
Fig.~\ref{fig:placement_sensor} shows a longitudinal B-mode image of the biceps brachii short head (acquired with an Ultrasonix Sonix 01 RP), on top of which we highlight the location of the A-mode \ac{US} channel and the sEMG electrodes.

\subsubsection{Measurement Protocol}
The measurement protocol involved three repetitions of \SI{10}{s} rest and \SI{10}{s} isometric voluntary contractions, corresponding to a duty cycle of \SI{50}{\%}.

\subsection{Mode of operation and firmware implementation}

\begin{figure}[b]
\vspace{-0.3cm}
\centerline{\includegraphics[width=1\columnwidth]{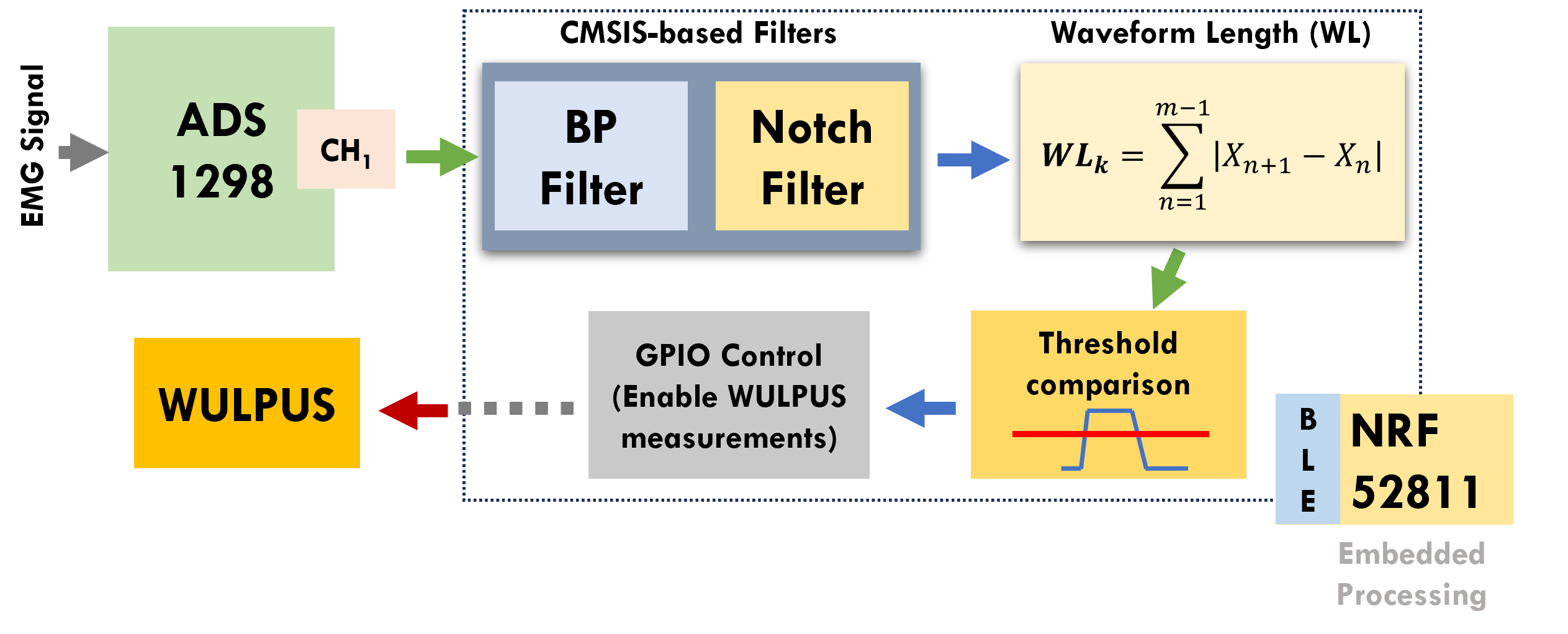}}
\vspace{-0.3cm}
\caption{Embedded sEMG Processing and US triggering}
\label{fig:embedded_processing}
\end{figure}
Our system operates by continuously processing sEMG data to activate the \ac{US} system when muscle activity is present. The embedded processing scheme is summarized in Fig.~\ref{fig:embedded_processing} and starts with BioGAP sampling a single differential sEMG channel through the ADS1298.
Sampled data are received by the nRF52811 \ac{MCU}, where it undergoes \ac{BP} filtering (using 3-taps IIR Butterworth) and notch filtering (using 3-taps IIR Butterworth). The filtering process is implemented with ARM CMSIS libraries, with guarantees low processing time and high energy efficiency\cite{cmsis}. To concentrate exclusively on muscle activation-related features, we employ a \ac{WL}-based envelope extraction method\cite{Kartsch2020}. The \ac{WL} extracts signal envelope by summing over the absolute difference between consecutive samples over a window of data (60 samples in this work). \ac{WL} is also low in computational complexity (w.r.t root-mean-square or time-frequency algorithms such as wavelet transform), which renders the algorithm very suitable for low-power applications.

Once the \ac{WL} signal surpasses a pre-established threshold (determined empirically as \SI{264}{mV}), the \ac{SoC} enables a dedicated \ac{GPIO} output, connected to the MSP430 \ac{MCU} of the \ac{US} sub-system, which reacts by initiating a continuous A-mode US sampling operating at \SI{50}{Hz}. The collected US data is subsequently transmitted to a PC through \ac{BLE} communication. Optionally, the synchronized raw or processed EMG data can also be transmitted through a separate \ac{BLE} channel.
During periods of low EMG activation, indicated by values below the threshold and resulting in a low \ac{GPIO} state, WULPUS enters a low-power mode. 

%% file: tables/ADS1298_configuration.tex
\begin{table}[t]
\begin{center}
\begin{threeparttable}[b]
\caption{Measurement configuration of the BioGAP AFE.}
\label{table:meas_config_ADS1298}
\scriptsize{
\begin{tabular}
{
p{0.8in}>
{\centering\arraybackslash}p{1in}>
{\centering\arraybackslash}p{1.7in}
}

\toprule[0.20em]

\textbf{Parameter} & \textbf{Value} \\
\midrule
Output data rate & \SI{500}{SPS}
\\
\midrule[0.12em]
\SI{-3}{dB} bandwidth  & \SI{131}{Hz}
\\
\midrule[0.12em]
PGA gain & \SI{6}{}
\\
\midrule[0.12em]
Resolution & \SI{24}{bit}
\\
\bottomrule[0.20em]

\end{tabular}}

\end{threeparttable}
\end{center}
\vspace{-6mm}
\end{table}

%% file: sections/experimental_verif_v2.tex
\section{Results and discussion}

\label{sec:results}
\vspace{-0.1cm}

\subsection{In vivo measurement}
\begin{figure}[htb]
\centerline{\includegraphics[width=0.89\columnwidth]{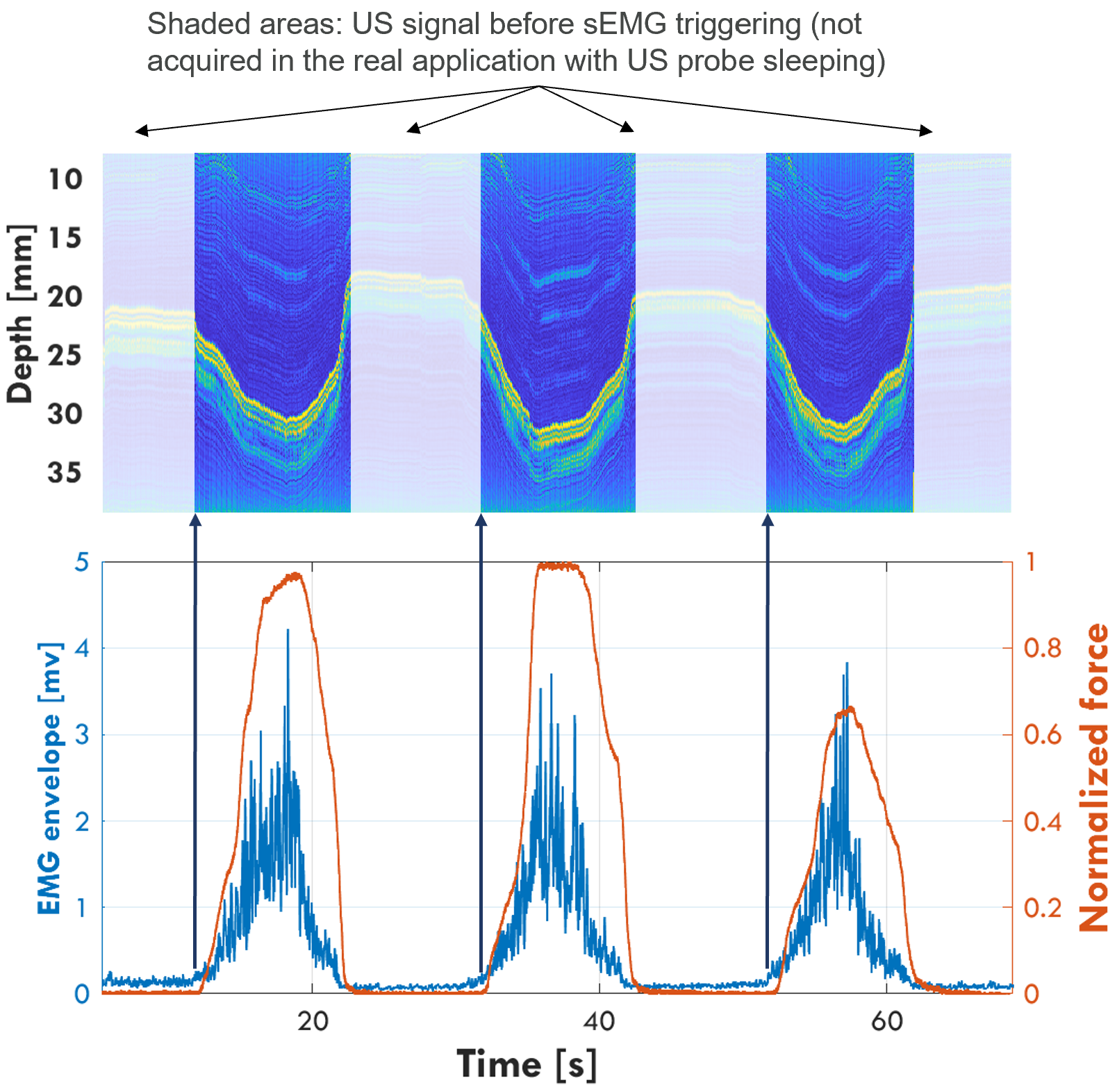}}
\vspace{-0.4cm}
\caption{Envelope of the measured sEMG signal and normalized force measurement with the \ac{US} activation threshold indicated in dark blue (bottom) and the measured M-mode \ac{US} signal (top). The partially transparent part of the \ac{US} signal is only shown for clearness and is not measured in the end-to-end application.}
\label{fig:EMG_US_measurement}
\end{figure}

Fig.~\ref{fig:EMG_US_measurement} (bottom) shows the force (normalized to a scale of 0 to 1 through division by the maximum value), as well as the sEMG envelope. Upon reaching the threshold of the sEMG envelope, the sEMG subsystem outputs the trigger (dark blue arrows), which activates the \ac{US} measurement with a \SI{50}{Hz} acquisition rate. Fig.~\ref{fig:EMG_US_measurement} (top) shows the corresponding M-mode \ac{US} measurement, where the three contractions can be seen clearly. The pronounced reflection shifting between approx. \SI{22}{-}\SI{30}{mm} with each contraction corresponds to the posterior boundary of the short head of the biceps brachii. For the sake of clarity, the semi-transparent part of the figure also shows the \ac{US} measurement when there is no sEMG activity. In the final application, the \ac{US} subsystem is in sleep mode during periods of low sEMG activity, and such data are not collected.

\subsection{Power measurements}
\begin{figure}[tb]
\centerline{\includegraphics[width=0.65\columnwidth]{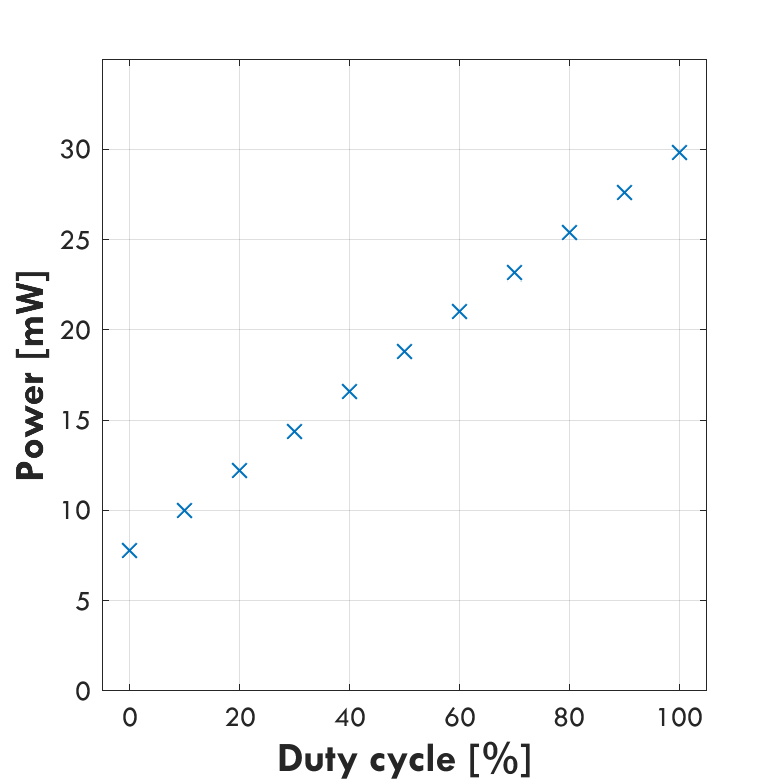}}
\vspace{-0.3cm}
\caption{Measured power consumption for different duty cycles of muscle contraction.}
\vspace{-0.4cm}
\label{fig:power_vs_duty_cycle}
\end{figure}

Figure~\ref{fig:power_vs_duty_cycle} presents the power consumption analysis of the entire system at various muscle contraction duty cycles. At the baseline of 0\% duty cycle (i.e., when BioGAP performs a single-channel sEMG measurement and onboard processing, while WULPUS is in sleep mode), the power consumption amounts to \SI{7.8}{mW}, corresponding to more than 6 days of battery life with a \SI{320}{mAh} battery. As the duty cycle increases, the activation of the \ac{US} system results in increasing power consumption. At 100\% duty cycle (i.e., both systems are active all the time), the total power consumption amounts to \SI{29.8}{mW}, resulting in a battery life of more than 1.5 days.

%% file: sections/conclusion.tex
\section{Conclusion}
\vspace{-0.1cm}
This work demonstrates a physiologically-informed triggering approach that integrates sEMG signals with wearable \ac{US} devices to improve the energy efficiency of heterogeneous biosignal acquisition systems. 
We demonstrate the use of sEMG signals as a trigger for initiating US measurements, as the electrical activity is responsible for the muscle contraction. The trigger generation is based on the enveloping and thresholding of sEMG signals and is faster than the electromechanical delay (\SI{30}{-}\SI{100}{ms}), thereby enabling to initiate US acquisitions before the displacement of muscle fascicles starts. This approach allows the \ac{US} system to be used only during the contraction phase, thereby boosting the battery lifetime. The system was implemented using WULPUS and BioGAP, two \ac{SoA} systems for wearable acquisition of US and ExG, respectively. Selectively using the US probe ensures higher energy efficiency with respect to continuous sampling. With \SI{200}{ms} long contractions at \SI{1}{Hz} contraction rate, the system consumes only \SI{12.2}{mW}, reducing the power envelope by \SI{59}{\%} compared to operating both sEMG and US continuously.
The proposed solution paves the way for the development of energy-efficient sensor-fusion solutions for sEMG and US, where US data are collected only when significant information is available.